# VeriODD: From YAML to SMT-LIB – Automating Verification of Operational Design Domains

1st Bassel Rafie *, 2nd Christian Schindler †, 3rd Andreas Rausch ‡,
*Institute for Software and Systems Engineering, Clausthal University of Technology*
Clausthal-Zellerfeld, Germany
*bassel.rafie@tu-clausthal.de , †christian.schindler@tu-clausthal.de , ‡andreas.rausch@tu-clausthal.de

*Abstract*—Operational Design Domains (ODDs) define the conditions under which an Automated Driving System (ADS) is allowed to operate, while Current Operational Domains (CODs) capture the actual runtime situation. Ensuring that a COD instance lies within the ODD is a crucial step in safety assurance. Today, ODD and COD specifications are frequently expressed in YAML to remain accessible for stakeholders, but such descriptions are not directly suitable for solver-based verification. Manual translation into formal languages such as SMT-LIB is slow and error-prone.

We present VeriODD, a tool that automates this translation. VeriODD uses ANTLR-based compiler technology to transform YAML-based ODD/COD specifications into both human-readable propositional logic, for lightweight review on a simple basis, and solver-ready SMT-LIB. The tool integrates with SMT solvers such as Z3 to provide automated consistency checks of ODD specifications and verification of COD conformance. A graphical user interface supports editing specifications, inspecting generated formulas, and performing verification with a single click.

VeriODD thereby closes the gap between stakeholder-friendly ODD/COD notations and formal verification, enabling scalable and automated assurance of operational boundaries in autonomous driving.

*Video demonstration: https://youtu.be/odRacNoL_Pk*
*Tool available at: https://github.com/BasselRafie/VeriODD*

## I. INTRODUCTION

**Motivation.** Operational Design Domains (ODDs) define the conditions under which an Automated Driving System (ADS) is intended to operate. In practice, these are essential for ensuring road safety and for regulatory approval of an ADS. At runtime, the Current Operational Domain (COD) reflects the actual driving situation and must be continuously checked against the ODD. This verification task is crucial for safety assurance in autonomous driving. As discussed in our earlier work [1], ADSs are assessed against ISO 26262 for functional safety, while ISO 21448 (SOTIF) addresses risks from functional insufficiencies that may arise when the system operates beyond its specified ODD.

**Problem description.** YAML has become a de facto standard for specifying ODDs and CODs in a semi-formal and human-readable way. Verification tools can be used to ensure that ODD specifications are consistent (i.e., free of contradictions) and to check whether a given COD is admissible during operation, but YAML descriptions cannot be directly processed by formal verification tools. To make this possible, the YAML descriptions must be translated into a solver-ready format such as SMT-LIB. Previous approaches relied on manual translation, which is slow, error-prone, and impractical for industrial-scale application.

**Research gap.** Although concepts for expressing ODDs in semi-formal formats and bridging them to formal verification exist, there is currently no tool that supports automated direct translation from YAML-based ODD/COD specifications into SMT-LIB and integrates solvers such as Z3 or CVC5. This gap hinders efficient consistency checking and runtime verification in safety-critical domains.

**Research challenge.** The key challenge is to automate the end-to-end process from YAML-based ODD/COD specifications to solver-ready logic, while preserving human readability, supporting modular reasoning, and enabling one-click verification of both ODD consistency and COD conformance.

**Basic approach.** To address this challenge, we developed VeriODD, a toolchain based on compiler technology with ANTLR4 grammars for ODD and COD. VeriODD automatically translates YAML-based ODD and COD descriptions into propositional logic for light review on a simple basis and into SMT-LIB for solver-based verification. A graphical user interface supports editing specifications, inspecting generated formulas, and performing automated verification with Z3.

**Contribution of the paper.** This paper presents VeriODD, a compiler-based tool that: (1) translates YAML ODD/COD specifications via ANTLR4, (2) supports translation to propositional logic and SMT-LIB, enabling lightweight inspection and formal verification, (3) integrates with Z3 SMT solver for automated ODD-consistency checking and COD-in-ODD verification, and (4) is evaluated on industrially motivated and OEM-provided ODD/COD examples, 145 golden unit tests, and runtime and scalability measurements.

**Structure of the rest of the paper.** The paper is organized as follows: Section II reviews related work and background, Section III details VeriODD's workflow, input language, and architecture, Section IV presents a running example and an evaluation, Section V concludes and outlines future work.

## II. RELATED WORK AND BACKGROUND

This section reviews related work in three directions. First, we examine how the concept of ODD is used in automotive and other industries such as avionics, railways, and maritime. Second, we treat ODD as part of the functional specification of safety-critical systems and therefore consider general methods for specification and verification.

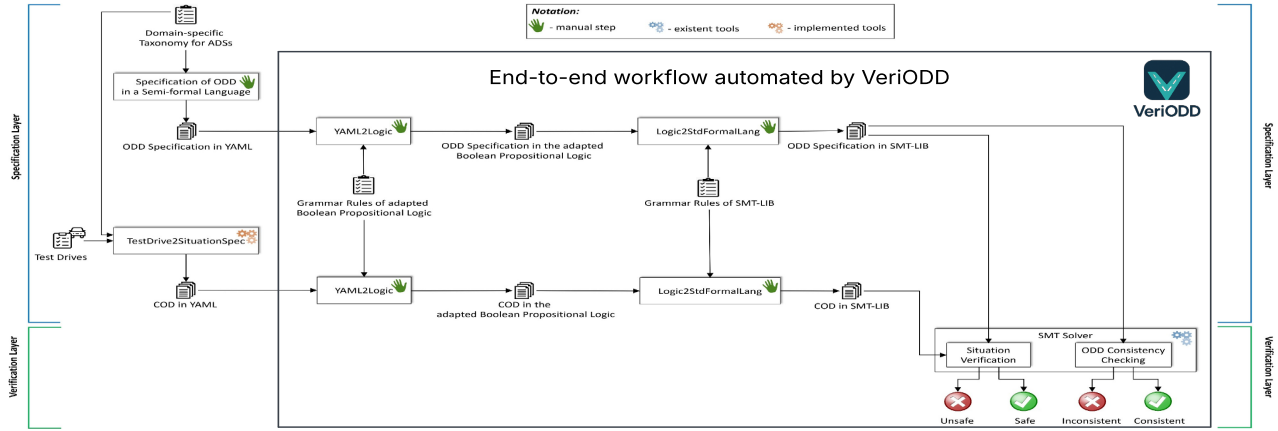

Fig. 1: Extended tool implementation: automating YAML-based ODD/COD translation into SMT-LIB for solver-based verification [1].

ODD was introduced in the automotive domain through standards such as SAE J3016 [2] and BSI 1883 [3], and by regulatory bodies like NHTSA [4]. It is commonly described via taxonomies (e.g., BSI 1883, Czarnecki [5]). Irvine et al. [6] proposed a human-readable ODD language, extended by Schwalb et al. [7] with SQL-based semantics. Sun et al. [8] introduced ODD acclimatization to validate safety in complex traffic, formalizing safety contracts in Signal Temporal Logic (STL), an extension of LTL. Other domains are also adopting ODD-like concepts as autonomy grows. Tonk and Boussif [9] proposed the operational envelope for maritime autonomous ships and the operational context for trains. Meng et al. [10] use ODD for scenario analysis in ATO, modeling components with UPPAAL timed automata and expressing requirements in its query language, a simplified TCTL. Cuartas et al. [11] apply UPPAAL with statistical model checking to verify requirements of a mechanical ventilator, addressing both symbolic and stochastic properties. Beyond model checking, formal verification also uses SAT- and SMT-based approaches, such as CVC5 [12] and Z3 [13]. Unlike model checking, which requires a system model and formalized properties (e.g., in LTL), SMT solvers verify satisfiability of logical formulas within background theories. Formal verification of ODD has been proposed and demonstrated by Aniculaesei et al. [1].

## III. WORKFLOW, INPUT LANGUAGE, AND ARCHITECTURE

**Workflow.** VeriODD automates the concept and methodological approach previously proposed and demonstrated in [1]. Figure 1 summarizes the steps now handled by VeriODD. Concretely, it replaces the manual workflow (hand-encoding YAML ODD/COD into propositional logic, transcribing the logic into SMT-LIB, assembling a solver script, and invoking the solver) with an end-to-end toolchain that generates SMT-LIB directly from YAML and runs Z3. VeriODD supports optional translation to propositional logic for light review on a simple basis, but it is not required for SMT-LIB generation.

**Input language.** The ODD/COD taxonomy was derived from OEM internal materials with [14] as the published baseline and refined toward the ASAM OpenODD YAML standard [15]. While the YAML structure and logical operators (e.g., *INCLUDE_AND*, *EXCLUDE_OR*, etc.) are fixed, the taxonomy itself remains flexible and can be adapted to different domains. Various examples can be found in the tool repository, and the grammars provide a detailed description of the taxonomy.

**Architecture.** Figure 2 presents an overview of VeriODD's architecture and processing pipeline, illustrating how the automated workflow is realized in practice. The tool employs two compiler pipelines implemented with ANTLR4 in Java. The core components are *ODD* and *COD* grammars, which contain input-language grammar rules that we defined and, using ANTLR4, generate a lexer and parser for recognizing each respective language. Each input is turned into a parse tree and traversed by code-generation visitors: *ODDVisitorSMTLIB* and *ODDVisitorPropositionalLogic*, as well as *CODVisitorSMTLIB* and *CODVisitorPropositionalLogic*. Each visitor walks the parse tree in a deterministic order and translates nodes into the destination language. After translation, the SMT-LIB outputs from ODD and COD are merged into a single verification script. At runtime, the user chooses the ODD module(s) to verify. COD conditions can be included or excluded. This enables two checks: (1) *ODD consistency checking*: asserting the chosen module(s) alone (without COD) to test internal satisfiability, and (2) *situation verification*: asserting the chosen module(s) together with COD to determine whether the concrete situation (COD) satisfies the ODD. Finally, the script is executed with the Z3 solver and the result is returned.

## IV. RUNNING EXAMPLE AND EVALUATION

### A. Running Example

We adopt as our running example the industrially motivated ODD and COD, specified in YAML for a parking-assistance system, from [1] (Listings 1-2). **Explanation:** The ODD consists of three modules: (1) *supported_parking_lot_conditions*: suitable conditions for operation (e.g., *parking_lot_length = 13*

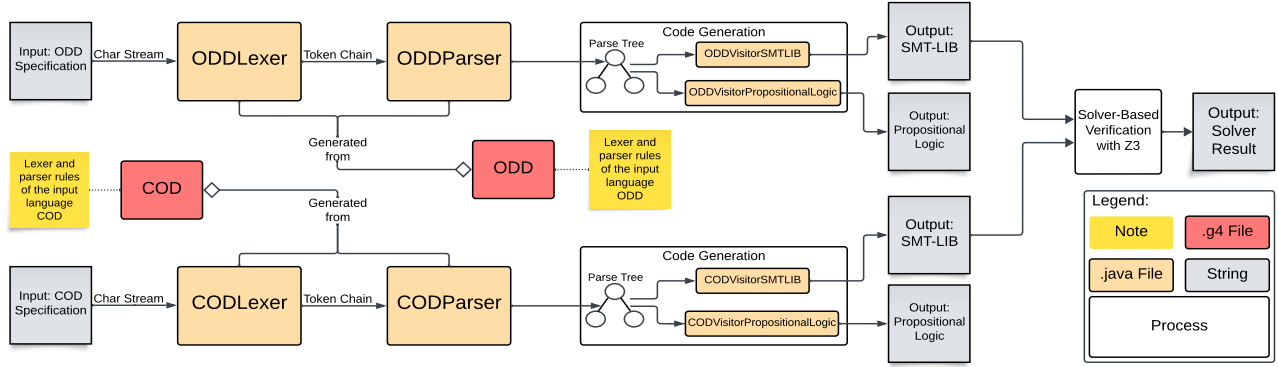

Fig. 2: Overview of VeriODD's architecture and processing pipeline [16].

∧ *is_curve = true*), (2) *unsupported_parking_lot_conditions*: unsuitable conditions for operation (e.g., *surface = puddle* ∧ *location = on_shoulder*), (3) *parking_lot_conditions*: top-level module that composes the first two modules, combining suitable and unsuitable operating conditions: (*suitable_conditions* ∧ ¬*unsuitable_conditions*). Listing 6 illustrates the identical YAML-ODD in propositional logic for more clarity. Listing 2 shows the conditions of the current driving situation (COD).

Listings 1–2 present the inputs (ODD, COD) and Listings 3-7 present the corresponding outputs of our compiler pipeline (SMT-LIB and propositional logic). The **six steps** of the running example are summarized as follows:

1) **ODD input:** The user provides the ODD (Listing 1). The ODD lexer and parser produce a parse tree.
2) **ODD code generation:** Visitors traverse the ODD tree and emit (i) ODD modules as SMT-LIB functions (Listing 3) and (ii) a propositional logic view (Listing 6).
3) **COD input:** The user provides a concrete driving situation (COD) in YAML (Listing 2). The COD lexer and parser produce a parse tree.
4) **COD code generation:** Visitors emit (i) COD conditions as assertions (Listing 4) and (ii) the propositional logic view (Listing 7).
5) **Assembly for verification:** The user selects one or more ODD modules, typically the top-level module to assert, and chooses whether to include or exclude the COD conditions. In this example, we assert the top-level module *parking_lot_conditions* and include the COD conditions. VeriODD then merges the ODD and COD SMT-LIB with the user's choice, and appends *(check-sat)* and, if requested, *(get-model)* (Listing 5).
6) **Solver run and result:** The assembled script is sent to Z3. The solver returns `SAT` or `UNSAT` (and, if requested, a model: variable assignments for a `SAT` case).

For this example, the solver returns `UNSAT` (the COD makes the *unsupported_parking_lot_conditions* module true, thus the top-level ODD module *parking_lot_conditions* is false). Selecting different modules, or multiple modules, enables targeted checks (e.g., asserting only *supported_parking_lot_conditions* returns `SAT`).

```yaml
supported_parking_lot_conditions:
    INCLUDE_AND:
        parking_lot_length: > 12 m
        is_curve: true

unsupported_parking_lot_conditions:
    INCLUDE_AND:
        surface:
            - puddle
            - snow_covered
        location:
            - on_shoulder
            - partly_on_subject_vehicle_lane

parking_lot_conditions:
    INCLUDE_AND:
        - supported_parking_lot_conditions
    EXCLUDE_OR:
        - unsupported_parking_lot_conditions
```

Listing 1: Example YAML ODD input for a Parking Assistance System [1].

```yaml
parking_lot_length: = 13
is_curve: true
surface: snow_covered
location: on_shoulder
```

Listing 2: Example YAML COD input for a real-time situation. [1].

```
(declare−const parking_lot_length Int)
(declare−const is_curve Bool)
(declare−const surface String)
(declare−const location String)

(define−fun supported_parking_lot_conditions () Bool
    (and (> parking_lot_length 12) is_curve))
(define−fun unsupported_parking_lot_conditions () Bool
    (and (or (= surface "puddle") (= surface "snow_covered"))
        (or (= location "on_shoulder")
            (= location "partly_on_subject_vehicle_lane"))))
(define−fun parking_lot_conditions () Bool
    (and supported_parking_lot_conditions
        (not unsupported_parking_lot_conditions)))
```

Listing 3: SMT-LIB output for the example ODD input.

```
1  (assert (= parking_lot_length 13))
2  (assert (= is_curve true))
3  (assert (= surface "snow_covered"))
4  (assert (= location "on_shoulder"))
```
Listing 4: SMT-LIB output for the example COD input.

```
1  (assert parking_lot_conditions)
2  (check-sat) ; + (get-model) if requested
```
Listing 5: Assertion of the user-selected module.

```
1  supported_parking_lot_conditions:=
2  [((parking_lot_length > 12) & is_curve)]
3
4  unsupported_parking_lot_conditions:=
5  [(((surface = puddle) | (surface = snow_covered)) &
6   ((location = on_shoulder) | (location =
7   partly_on_subject_vehicle_lane)))]
8
9  parking_lot_conditions:=
10 [[((parking_lot_length > 12) & is_curve)] &
11  (![(((surface = puddle) | (surface = snow_covered)) &
12  ((location = on_shoulder) | (location =
13  partly_on_subject_vehicle_lane)))])]
```
Listing 6: Propositional logic output for the example ODD input.

```
1  parking_lot_length = 13
2  is_curve = true
3  surface = snow_covered
4  location = on_shoulder
```
Listing 7: Propositional logic output for the example COD input.

### B. Evaluation

We evaluated VeriODD on **three criteria**: (1) applicability to industrial ODDs/CODs, (2) semantically consistent generation of propositional logic and SMT-LIB encodings, and (3) runtime behavior. **Applicability.** On the industrially motivated example (Listings 1–2), VeriODD runs end-to-end: it parses YAML, generates SMT-LIB and propositional logic outputs, supports module assertions, and queries Z3, demonstrating practical applicability on industrial-style specs. It was also tested with OEM materials (not shareable due to NDAs) and functioned as intended. **Correctness.** We implemented 145 golden unit tests covering all four translators ODD/COD to propositional logic and SMT-LIB, each asserting equality with the expected outputs. **Runtime.** We validated 10–5000 CODs against a 6-variable ODD (Listing 1) and a 1000-variable ODD, creating a fresh solver instance per COD. End-to-end timings (translation, script assembly, solver invocation) grow approximately linearly with the number of CODs (Figure 3).

## V. CONCLUSION AND FUTURE WORK

VeriODD is an ANTLR-based tool that compiles YAML ODD/COD into (1) propositional logic for lightweight review and (2) SMT-LIB for solver-based verification in Z3. It supports module-level reasoning by allowing users to assert one or more ODD modules for consistency (without COD) or COD conformance (with COD), and it provides one-click assembly and execution of the combined ODD/COD script. Future work includes fully aligning VeriODD with the latest ASAM OpenODD YAML standard [15] and providing a public API.

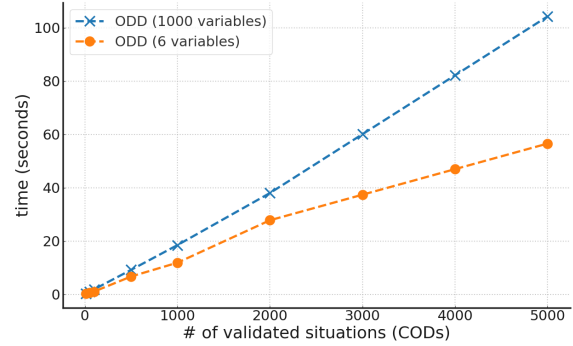

Fig. 3: End-to-end runtime of checking different amounts of CODs against a small and a large ODD.

ACKNOWLEDGMENT

This acknowledgment goes to generative AI namely ChatGPT for assisting in reviewing its language, grammar, and restructure some paragraphs.